\documentclass[12pt] {article}
\pdfoutput=1
\usepackage[hyphens,spaces,obeyspaces]{url}
\usepackage{setspace}

\usepackage{epsfig}
\usepackage{amssymb}
\usepackage{amsmath}
\usepackage{multirow}
\usepackage{setspace}
\usepackage{datetime}
\usepackage{amsmath}
\usepackage{algorithm}
\usepackage{algpseudocode}
\usepackage{graphicx}
\usepackage{subfig} 
\usepackage{lipsum}
\usepackage{acronym}
\usepackage{array}
\usepackage{graphicx}
\usepackage{cite}
\usepackage{graphicx}
\usepackage{epstopdf}
\usepackage{caption}
\usepackage{balance}
\usepackage{fancybox}
\usepackage{colortbl}
\usepackage{array}
\usepackage{mystyle}
\usepackage{multirow}
\usepackage[obeyspaces,hyphens,spaces]{url}
\usepackage{threeparttable}
\usepackage[english]{babel}
\usepackage{longtable}
\usepackage{listings, lstautogobble}
\usepackage{booktabs}
\usepackage{threeparttable}
\usepackage[table,xcdraw]{xcolor}

\usepackage{amsthm}
\theoremstyle{definition}
\AtBeginEnvironment{thm}{\footnotesize}
\AtBeginEnvironment{defn}{\footnotesize}
\renewcommand{\baselinestretch}{1}

\providecommand{\keywords}[1]{\textbf{\textit{Keywords---}} #1}

{\begin{list}{}%
         {\setlength{\leftmargin}{#1}}%
         \item[]%
}
{\end{list}}

\newdateformat{monthyeardate}{%
  \monthname[\THEMONTH], \THEYEAR}
\begin{document}
\title{\Large{\textbf{Error Analysis of Approximate Array Multipliers} }}
\author{
Mahmoud~Masadeh$^{1}$, Osman~Hasan$^{1,2}$, and Sofi\`ene~Tahar$^{1}$ \vspace*{2em}\\
$^{1}$Department of Electrical and Computer Engineering,\\
Concordia University, Montr\'eal, Canada  \\
\{m\_masa,~o\_hasan,~tahar\}@ece.concordia.ca \vspace*{2em}\\  \\
$^{2}$Electrical Engineering and Computer Science,\\
National University of Science and Technology, Islamabad, Pakistan \\ \\  \\ \\ \\
\textbf{TECHNICAL REPORT}\\ \\ 
\date{August 2019}
}
\maketitle

\newpage
\begin{abstract}
Approximate computing is a nascent energy-efficient computing paradigm suitable for error-tolerant applications. However, the value of approximation error depends on the applied inputs where individual output error may reach intolerable level while the average output error is acceptable. Thus, it is critical to show the response of approximate design for various applied inputs where understanding the interdependence between the inputs and the approximation error could be utilized to control the output quality. In this report, we exhaustively analyze the accuracy of 20 different designs of 8-bit approximate array multipliers. We designed these multipliers based on various configurations including the \textit{type} of approximable component and how much of the result to approximate, i.e., approximation \textit{degree}. Accuracy analysis shows a strong correlation between the applied inputs and the magnitude of the observed error, i.e., error distance, normalized error distance and peak-to-signal-noise ration. We may utilize this input-dependency of approximation error, in controlling the quality of approximate computing by eliminating large magnitude errors and improving the quality of the final results.

\end{abstract}

\newpage
\tableofcontents
\newpage
\section{Introduction}

Approximate Computing (AxC) known as best-effort, inexact and tolerable computing is gaining a great attention due to the unprecedented benefits. This emerging computing paradigm is suitable for processing the gigantic data produced in the era of ``IoT", where the world produces 2.5 quintillion bytes of data per day \cite{IoT1}. Many modern applications including web searching, image processing, machine learning, and computer vision show a tendency to accept imperfect results. Approximate computing is a viable method of exploiting error tolerance in such applications to trade accuracy for energy
savings and performance improvements.

Multiple approximation techniques have been proposed at different abstraction layers of the computing stack \cite{Rehman}. At the software layer, techniques like loop unrolling and code approximation are commonly employed while, at the hardware layer, techniques like approximation of the functional units, e.g., adders \cite{AAD1}, dividers \cite{Div3} and multipliers \cite{Suganthi}, are commonly used. A detailed description of approximation techniques and their computing layers can be found in \cite{S1}.

In \cite{Masadeh}, we designed a set of 8-bit and 16-bit approximate multipliers with acceptable average output quality and reduced area, delay and power consumption. These multipliers are designed based on three design settings: (1) the type of approximate full adder (FA) used to construct the multiplier; (2) the architecture of the multiplier, i.e., array or tree; and (3) the placement of sub-modules of approximate and exact multipliers in the target multiplier module. We were able to design approximate multipliers, which are suitable to applications with intrinsic error resiliency. We used these designs in an image processing application and obtained interesting results, thus they are applicable to other domains. However, utilizing a specific approximate design for the whole range of the applied inputs could introduce a large magnitude error for specific inputs where such error is undesirable. In this report, we investigate the relationship between the range of design input values and their associated approximation errors in order to identify the individual inputs which cause a large approximation error. For that, we rely on a library of approximate multipliers which encompasses 20 different designs of 8-bit approximate array multipliers. We were able to clearly show the dependency of various approximation metrics on the input data.

The rest of this report is organized as follows: Section \ref{sec:ApproxLib} describes our library of approximate multipliers with 20 designs. The analysis of various error metrics of the approximate designs is explained in Section \ref{sec:AccuracyAnalysis}. Finally, some conclusions are drawn in Section \ref{sec:conclusion}.

\section{Library of Approximate Multipliers	} \label{sec:ApproxLib}

Aiming to design 8-bit and 16-bit approximate multipliers, in \cite{Masadeh} we investigated five approximate mirror adders (AMA) \cite{Vaibhav}, named AMA1--AMA5, three approximate XOR/XNOR based full adders (AXA) \cite{XORFA}, named AXA1--AXA3 and three inexact adders (InXA) \cite{InXA}, named InXA1--InXA3. These FAs constitute the basic building blocks to design array multipliers. Moreover, we utilized these 11 approximate FAs to construct approximate compressors, i.e., 4-to-3, 5-to-3, 6-to-3, 7-to-3 and 8-to-4, which are the basic blocks for tree multipliers. The approximable building blocks, i.e., FAs and compressors, were used to construct approximate multipliers, i.e., array or tree architecture, where the final result has a variable number of approximable bits, i.e., 7-bits, 8-bits, 9-bits and 16-bits.  


In this report, in order to investigate the relationship between the applied inputs to an approximate design and the associated error, we select 8-bit approximate array multiplier designs based on the five approximate mirror adders (AMA1, AMA2, AMA3, AMA4 and AMA5), where the final result has 7-bits, 8-bits, 9-bits and 16-bits of the results being approximated. We call such designs our ``Library Approximate Multipliers".

Generally, an approximate design with N knobs, i.e., \textit{K1}, \textit{K2},..., and \textit{KN}, will have $\left\vert{K1}\right\vert$x$\left\vert{K2}\right\vert$x...x$\left\vert{KN}\right\vert$ different design settings. The 8-bit approximate multipliers of our library have two knobs; \textit{K1} which represents the \textit{type} of the approximate block used to build the approximate design, and \textit{K2} which is the approximation \textit{degree} of the design. Thus, K1= $\lbrace AMA1, AMA2, AMA3, AMA4, AMA5 \rbrace$, i.e., chosen from the low power approximate full adders \cite{Vaibhav} and K2=$\lbrace D1, D2, D3, D4 \rbrace$, where $D1$ has 7 bits approximated out of 16-bit result, while $D2$, $D3$, and $D4$ have 8, 9, and 16 approximate bits, respectively. We have $\left\vert{K1}\right\vert$x$\left\vert{K2}\right\vert$=$5$x$4$=$20$ different settings (as shown in Table \ref{tab:Lib}) with enhanced design characteristics, i.e., reduced power consumption and execution time, compared to the exact design \cite{Masadeh}.



\begin{table}[t!]
\centering
\caption{Library of approximate multipliers with 20 configurations based on \textit{Degree} and \textit{Type} knobs}
\label{tab:Lib}
\begin{tabular}{|l|l|c|c|c|c|}
\hline
\multicolumn{2}{|c|}{\multirow{2}{*}{\textbf{\begin{tabular}[c]{@{}c@{}}Approximate \\ Design\end{tabular}}}} & \multicolumn{4}{c|}{\textbf{Degree}} \\ \cline{3-6} 

\multicolumn{2}{|c|}{} & \multicolumn{1}{c|}{\textbf{D1}} & \multicolumn{1}{c|}{\textbf{D2}} & \multicolumn{1}{c|}{\textbf{D3}} & \multicolumn{1}{c|}{\textbf{D4}} \\ \hline

\multirow{5}{*}{\textbf{Type}} & \textbf{AMA1} & \textit{Design1} & \textit{Design2} & \textit{Design3} & \textit{Design4} \\ \cline{2-6} 
 & \textbf{AMA2} & \textit{Design5} & \textit{Design6} & \textit{Design7} & \textit{Design8} \\ \cline{2-6} 
 
 & \textbf{AMA3} & \textit{Design9} & \textit{Design10} & \textit{Design11} & \textit{Design12} \\ \cline{2-6} 
 
 & \textbf{AMA4} & \textit{Design13} & \textit{Design14} & \textit{Design15} & \textit{Design16} \\ \cline{2-6} 
 
 & \textbf{AMA5} & \textit{Design17} & \textit{Design18} & \textit{Design19} & \textit{Design20} \\ \hline
\end{tabular}
\end{table}
\section{Error Analysis} \label{sec:AccuracyAnalysis}

In this section, we are going to analyze the accuracy of the approximate library, i.e., 20 designs of 8-bit approximate multipliers, based on various error metrics. Then, the accuracy of the approximate designs is analyzed to verify its input-dependency which is an essential consideration in approximate computing. Based on that, we identify the accuracy metrics that are the most suitable to be used in different situations/applications.


\subsection{Accuracy Metrics} \label{sec:AccMetrics}

Approximation introduced accuracy as a new design metric for digital designs. There are several application dependent \textit{error metrics} used in AxC to quantify approximation errors and evaluate design accuracy \cite{InXA}. For example, considering an approximate design with two inputs, i.e., $X$ and $Y$, of \textit{n}-bit each, where the exact result is ($P$) and the approximate result is ($P'$), these error metrics include:

\begin{itemize}  
\item Error Rate (ER): Also called error probability, is the percentage of erroneous outputs among all outputs.
\item Error Distance (ED): The arithmetic difference between the exact output and the approximate output for a given input. ED can be given by:
 \begin{equation}
 ED= |P - P'|
 \end{equation}
\item Mean Error Distance (MED): The average of ED values for a set of outputs obtained by applying a set of inputs. MED is an effective metric for measuring the implementation accuracy of a multiple-bit circuit design, which is obtained as:
\begin{equation}
MED = \frac{1}{2^{2n}} {\sum_{i=1}^{2^{2n}} |ED_i| }
\end{equation}

\item Normalized Error Distance (NED): The normalization of MED by the maximum result that an exact design can have ($P_{max}$). NED is an invariant metric independent of the size of the circuit, therefore, it is used for comparing circuits of different sizes, and it is expressed as:

\begin{equation}
NED = \frac{MED}{P_{max}} 
\label{equ:NED}
\end{equation}


\item Relative Error Distance (RED): The ratio of ED to the accurate output, given by:  
 \begin{equation}
RED= \frac{ED}{P} = \frac{|P - P'|}{P}
\label{equ:RED}
 \end{equation}

\item Mean Relative Error Distance (MRED): The average value of all possible relative error distances (RED):

 \begin{equation}
MRED = \frac{1}{2^{2n}} {\sum_{i=1}^{2^{2n}} |RED_i| }
\label{equ:MRED}
 \end{equation}

\item Mean Square Error (MSE): It is defined as the average of the squared ED values:

 \begin{equation}
MSE = \frac{1}{2^{2n}} {\sum_{i=1}^{2^{2n}} {|P_i - P'_i|}^2 } =  \frac{1}{2^{2n}} {\sum_{i=1}^{2^{2n}} {|ED_i|}^2 }
\label{equ:MSE}
 \end{equation} 
 
\item Peak Signal-to-Noise Ratio (PSNR): The peak signal-to-noise ratio is a fidelity metric used to measure the quality of the output images, and given by:
 \begin{equation}
PSNR = 10*log_{10}(\frac{255^2}{MSE}) 
\label{equ:PSNR}
 \end{equation} 
 \end{itemize}

\begin{table*}[ht!]
\centering
\caption{Accuracy metrics for the library of approximate multipliers}
\label{tbl:ErrorMetrics}
\resizebox{0.99\textwidth}{!}{%
\begin{tabular}{|c|c|c|c|c|c|c|c|}
\hline
\textbf{\begin{tabular}[c]{@{}c@{}}Approx\\  Degree\end{tabular}} & \textbf{FA Type} & \textbf{ER} & \textbf{MED} & \textbf{NED} & \textbf{MRED} & \textbf{MSE} & \textbf{PSNR}  \\ \hline \hline

\multirow{5}{*}{\textbf{D1}} & \textbf{AMA1} & 0.931 & 102 & 0.0165 & 0.0380 & 1.69E+04 & 39.35    \\ \cline{2-8}  
& \textbf{AMA2} & 0.978 & 101 & 0.0213 & 1.0447 & 1.44E+04 & 39.97 \\ \cline{2-8}   
& \textbf{AMA3} & 0.996 & 200 & 0.0416 & 2.8055 & 4.76E+04 & 34.78 \\ \cline{2-8}   
& \textbf{AMA4} & 0.932 & 51 & 0.0083 & 0.0189 & 4.11E+03 & 45.58 \\ \cline{2-8}    
& \textbf{AMA5} & 0.870 & 44 & 0.0069 & 0.0148 & 3.14E+03 & 46.80 \\ \cline{2-8}   
 
\hline  \hline
\multirow{5}{*}{\textbf{D2}} & \textbf{AMA1} & 0.962 & 246 & 0.0366 & 0.0823 & 9.61E+04 & 32.10 \\ \cline{2-8}     
& \textbf{AMA2} & 0.990 & 227 & 0.0515 & 2.1470 & 7.23E+04 & 33.26 \\ \cline{2-8}   
& \textbf{AMA3} & 0.999 & 474 & 0.1077 & 7.0425 & 2.68E+05 & 27.65 \\ \cline{2-8}   
& \textbf{AMA4} & 0.963 & 107 & 0.0164 & 0.0350 & 1.79E+04 & 39.15 \\ \cline{2-8}   
&\textbf{AMA5} & 0.922 & 93 & 0.0138 & 0.0280 & 1.38E+04 & 40.32 \\ \cline{2-8}  

 \hline \hline
 
\multirow{5}{*}{\textbf{D3}} & \textbf{AMA1} & 0.977 & 538 & 0.0758 & 0.1679 & 4.53E+05 & 25.80  \\ \cline{2-8} 
& \textbf{AMA2} & 0.996 & 464 & 0.0976 & 4.2929 & 2.97E+05 & 27.41 \\ \cline{2-8} 
& \textbf{AMA3} & 1.000 & 1010 & 0.2280 & 13.4428 & 1.17E+06 & 21.75 \\ \cline{2-8}
& \textbf{AMA4} & 0.977 & 185 & 0.0286 & 0.0582 & 5.27E+04 & 34.30 \\ \cline{2-8}
& \textbf{AMA5} & 0.952 & 185 & 0.0261 & 0.0488 & 5.34E+04 & 34.56 \\ \cline{2-8}
  \hline  \hline
\multirow{5}{*}{\textbf{D4}} & \textbf{AMA1} & 0.9882 & 13162 & 1.1862 & 2.1 & 3.27E+08 & 5.54  \\ \cline{2-8}  
&\textbf{AMA2} & 0.9998 & 16902 & 5.0033 & 272.6 & 4.00E+08 & 8.18 \\ \cline{2-8}  
& \textbf{AMA3} & 0.9999 & 17211 & 3.9303 & 180.7 & 4.13E+08 & 7.36 \\ \cline{2-8} 
& \textbf{AMA4} & 0.9882 & 6334 & 0.4705 & 0.6039 & 7.91E+07 & 10.38 \\ \cline{2-8} 
& \textbf{AMA5} & 0.9825 & 8096 & 0.5309 & 0.6642 & 1.17E+08 & 8.85 \\ \hline 

\end{tabular}}
\end{table*}

\begin{figure*}[ht!]
\centering
\includegraphics[width=\textwidth]{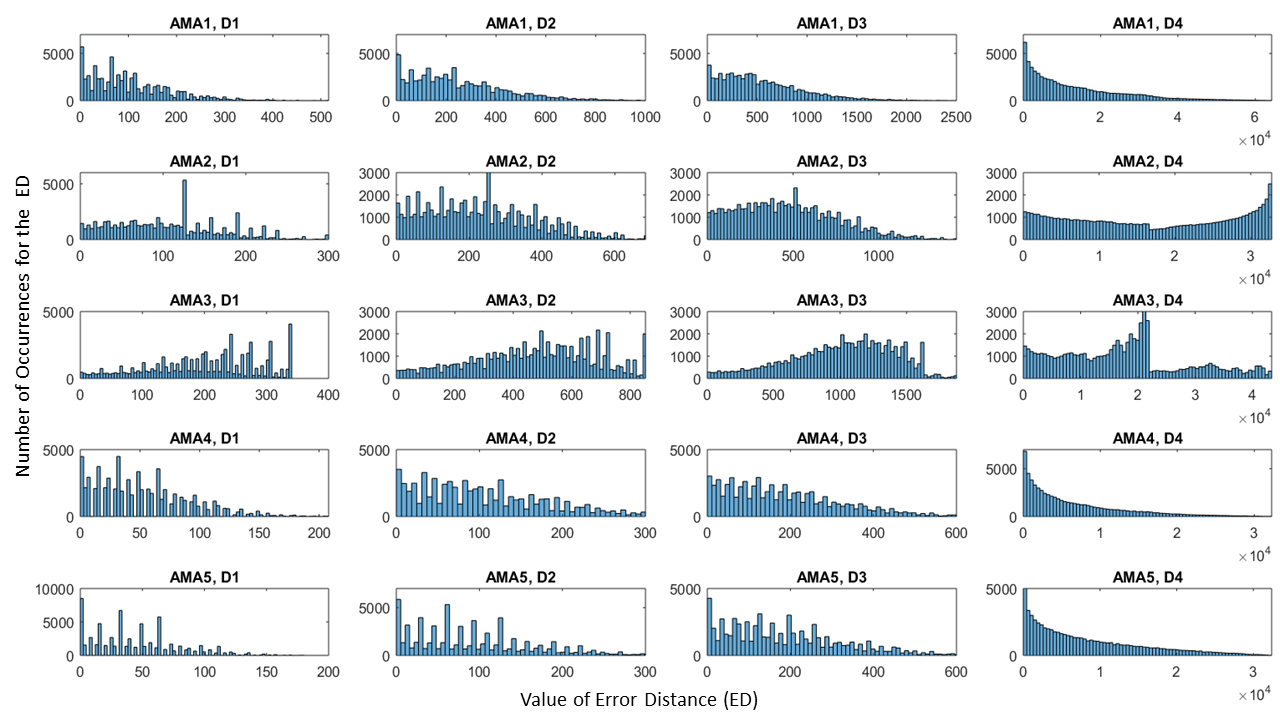} 
\caption{Histogram distribution of error distance (ED) for the library of approximate multipliers}
\label{fig:ED} 
\end{figure*}

Next, we present an error analysis for our library of approximate designs in order to observe the dependency of approximation error on the applied inputs. Such observations could be leveraged to control approximation quality by designing an input-aware adaptive approximate computing. Table \ref{tbl:ErrorMetrics} summarizes various error metrics, i.e., ER, MED, NED, MRED and PSNR, that we obtained based on an exhaustive simulation for 20 approximate designs. The shown values are averaged over the full range of possible inputs, and provide useful insights about the accuracy of designs. In the sequel, we describe the ED, NED and PSNR error metrics one by one in detail.

\subsection{Analysis of ED} \label{sec:ED}

Figure \ref{fig:ED} shows the histogram distribution of the error distance (ED) for the approximate multipliers given in Table \ref{tab:Lib}. The ED for the designs based on \textit{AMA1} with D1, D2, D3 and D4 have the average of 102, 246, 538 and 13162, respectively. While the ED for the designs based on \textit{AMA2} with D1, D2, D3 and D4 have the average of 101, 227, 464 and 16902, respectively. Designs based on \textit{AMA3} with D1, D2, D3 and D4 have the average ED of 200, 474, 1010 and 17211, respectively. However, the designs based on \textit{AMA4} with D1, D2, D3 and D4 have the average ED of 51, 107, 185 and 6334, respectively, which is quite low. \textit{AMA5 based designs with D1, D2, D3 and D4 have the lowest average of ED which is 44, 93, 185 and 8096, respectively}.

We notice that the error distance varies for different inputs, for example, \textit{Design1} (top-left corner of Figure \ref{fig:ED}) shows that the ED varies from 0 to 518 with an average of 102. If the value of ED was input-independent, then it would constant for all  input values. Such input-dependency is also present in the other designs, i.e., \textit{Design2}--\textit{Design20}. Moreover, we notice that the variation in ED based on approximation degree for a specific type (represented horizontally in Figure \ref{fig:ED}) is more evident than the variation between different types for a specific degree (represented vertically in Figure \ref{fig:ED}), i.e., ED is correlated with the degree more than with the type. Clearly, the ED increases (almost doubles) while increasing the approximation degree for any design type. Ideally, for every input, there exist a specific design (among the 20 designs) with a minimal error distance (ED) which is the most suitable to be used in error-tolerant applications. However, changing the most suitable design for every input is impractical due to the associated overhead. 

\subsection{Analysis of NED} \label{sec:NED}
\begin{figure*}[th!]
\centering
\includegraphics[width=\textwidth]{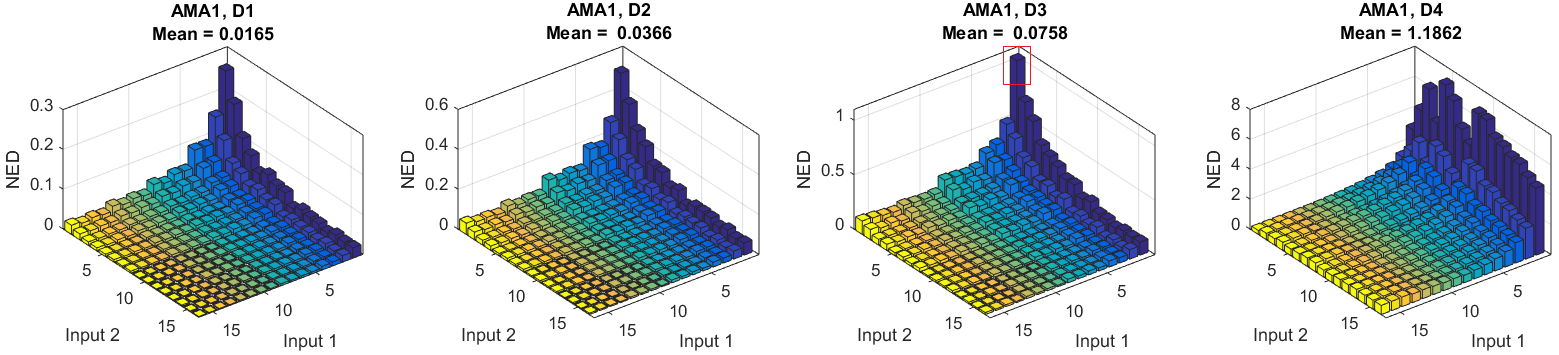} 
\caption{NED for approximate multiplier based on \textit{AMA1} type and D1, D2, D3 and D4 approximation degrees}
\label{fig:NED_1} 
\vspace{-0.1cm}
\end{figure*}

\begin{figure*}[th!]
\centering
\includegraphics[width=\textwidth]{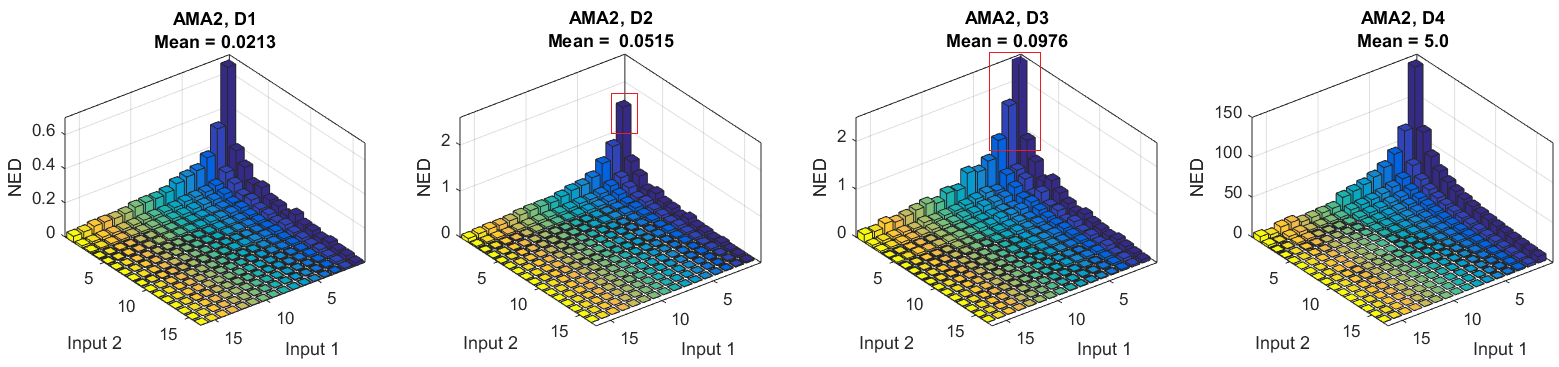} 
\caption{NED for approximate Multiplier based on \textit{AMA2} type and D1, D2, D3 and D4 approximation degrees}
\label{fig:NED_2} 
\vspace{-0.1cm}
\end{figure*}

\begin{figure*}[th!]
\centering
\includegraphics[width=\textwidth]{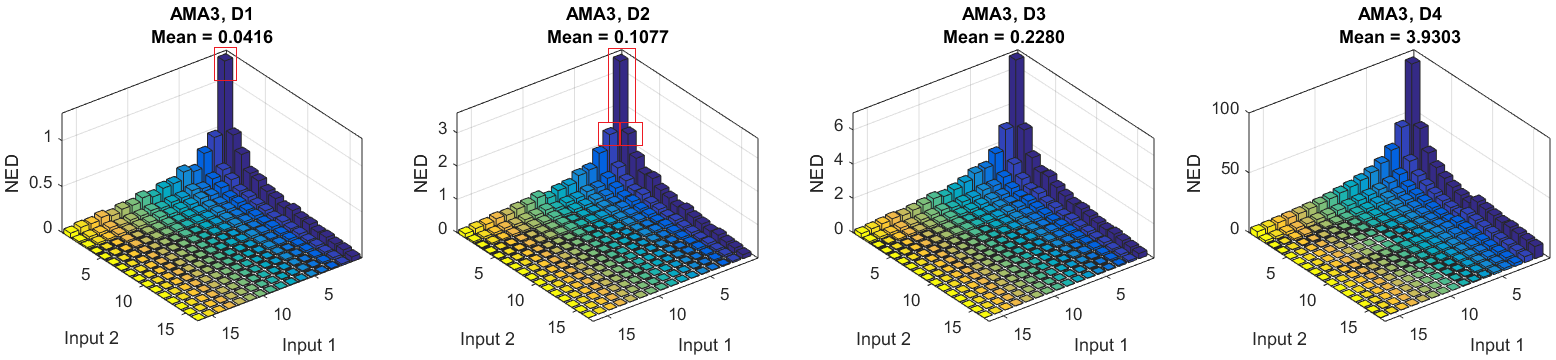} 
\caption{NED for approximate Multiplier based on \textit{AMA3} type and D1, D2, D3 and D4 approximation degrees}
\label{fig:NED_3} 
\vspace{-0.1cm}
\end{figure*}

\begin{figure*}[h!]
\centering
\includegraphics[width=\textwidth]{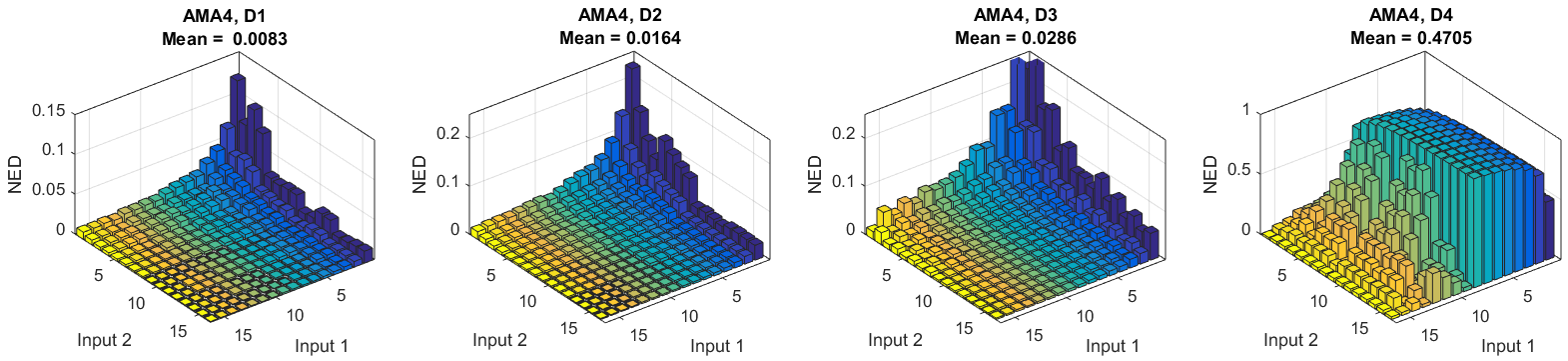} 
\caption{NED for approximate Multiplier based on \textit{AMA4} type and D1, D2, D3 and D4 approximation degrees}
\label{fig:NED_4} 
\vspace{-0.1cm}
\end{figure*}

\begin{figure*}[th!]
\centering
\includegraphics[width=\textwidth]{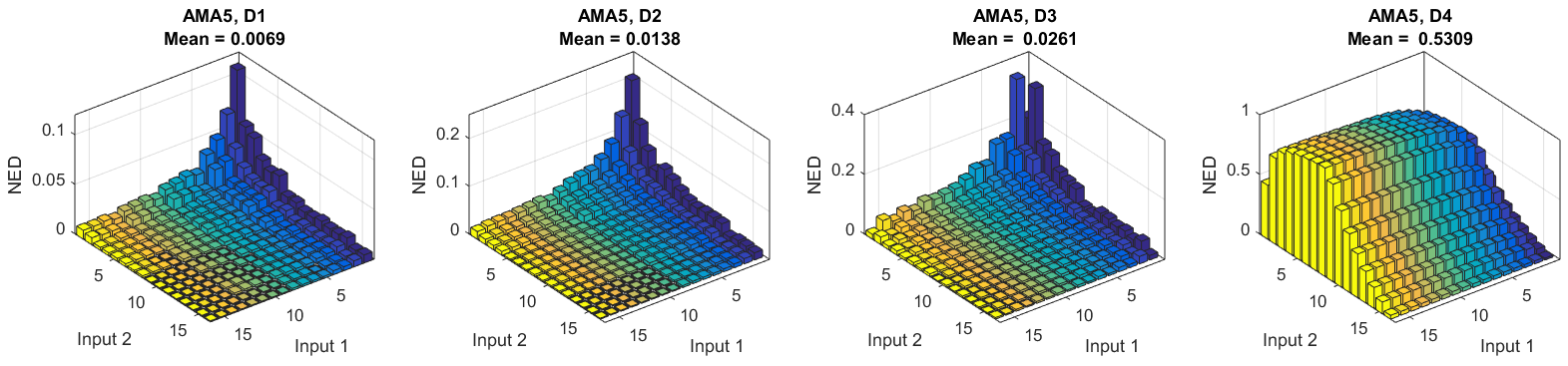} 
\caption{NED for approximate Multiplier based on \textit{AMA5} type and D1, D2, D3 and D4 approximation degrees}
\label{fig:NED_5} 
\vspace{-0.1cm}
\end{figure*}

As explained previously in Equation \ref{equ:NED}, NED is an invariant metric, which is independent of the size of the approximate design. Figures \ref{fig:NED_1}--\ref{fig:NED_5} show the NED for different approximate multipliers, utilizing \textit{AMA1}--\textit{AMA5} types, respectively. Each figure depicts a graphical representation for four approximation degrees, i.e., \textit{D1}, \textit{D2}, \textit{D3} and \textit{D4}. Since NED should be computed over a range of inputs rather than a single input, we clustered every 16 consecutive inputs as a group to be able to compute the average of NED over it. 

Each bar -from the 256 bars- in these figures represents the average error (NED) of the approximate design based on 16 consecutive values of the first input, i.e., \textit{Input1}, and 16 consecutive values of the second input, i.e., \textit{Input2}. A  high error value, e.g., $NED \geq 20\%$, indicates an unacceptable error in most applications. These designs are thus not selected. Now, for clarification purposes, we consider NED with a value of $\leq 100\%$ as our threshold.

Figure \ref{fig:NED_1} shows the NED for the designs based on \textit{AMA1}, where the design based on \textit{D1} has a NED average and maximum NED of 1.7\% and 26.2\%, respectively. Similarly, the \textit{D2} based design has an average and maximum NED values of 3.7\% and 51\%, respectively, while the \textit{D3} based design has an average of 7.6\%. However, one out of the 256 clusters has a value greater than our specified threshold, i.e., $NED \leq100\%$. Other clusters have values less than 71.6\%. Whereas, for the \textit{D4} based design, 79 clusters out of 256 have a $NED>100\%$, with an average of 118.6\%.

Figure \ref{fig:NED_2} shows the NED for designs based on the \textit{AMA2} type. The design based on \textit{D1} has an average of 2.1\% with a maximum value of 68\%, while the \textit{D2} based design has an average of 5.1\% with one cluster having a value of 164\%, and the other 255 clusters have a maximum value of 87.8\%. Similarly, the \textit{D3} based design has an average of 9.8\% with 4 clusters having a value $ \geq 100\%$, and the remaining 252 clusters have a maximum value of 70.3\%.

The NED for designs based on \textit{AMA3} type are shown in Figure \ref{fig:NED_3}. The \textit{D1} based design has an average of 4.1\% with one cluster having a value of 128\%, while the other 255 clusters have a maximum value of 49.3\%. Similarly, the \textit{D2} based design has an average of 10.8\% with 3 clusters having a value $>100\%$, and the other 253 clusters have a maximum value of 99\%. The \textit{D3} based design has a large average of NED, which is around 22.8\% and many clusters have $>100\%$ NED values. Moreover, the \textit{D4} based design has a large error for most of the inputs.

Figures \ref{fig:NED_4} and \ref{fig:NED_5} show the NED for designs based on \textit{AMA4} and \textit{AMA5}, respectively. As depicted graphically, all designs have a low value of NED. Figure \ref{fig:NED_4} shows that the maximum values of NED for \textit{AMA4} based designs are 12.5\%, 23\%, 37\% and 91\% for \textit{D1}--\textit{D4} based designs, respectively. Similarly, as shown in Figure \ref{fig:NED_5}, the maximum values of NED for \textit{AMA5} based designs are 11\%, 21\%, 35\% and 91\% for \textit{D1}--\textit{D4} based designs, respectively. As we noticed, for a given error threshold, e.g., $NED \leq 100\%$, the average is acceptable for most of the input clusters. However, there are some cases where it has a large error value and such cases should be identified.

\subsection{Analysis of PSNR} \label{sec:PSNR}

\begin{figure*}[th!]
\centering
\includegraphics[width=\textwidth, height=0.17\textheight]{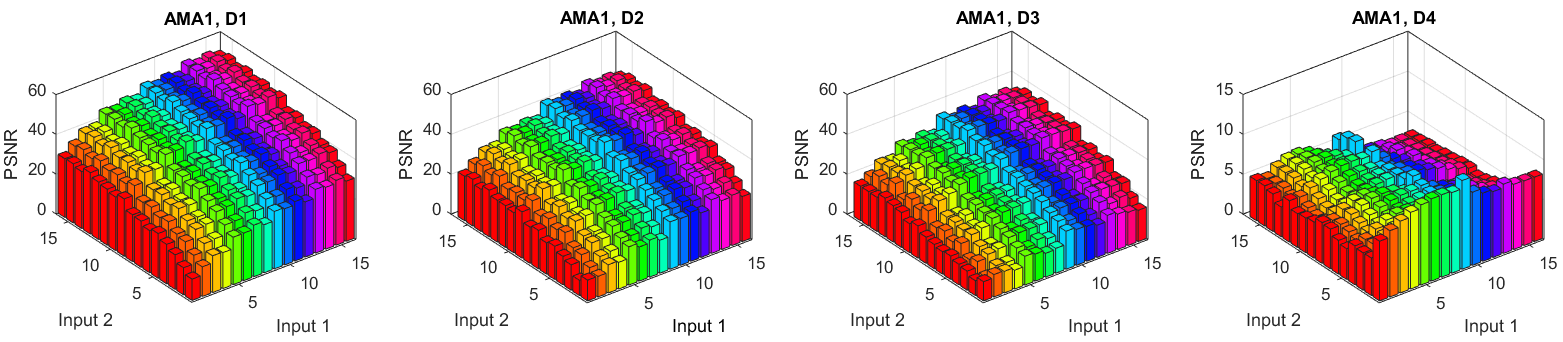} 
\caption{PSNR for approximate multiplier based on \textit{AMA1} type and D1, D2, D3 and D4 approximation degrees}
\label{fig:PSNR_1} 
\vspace{-0.1cm}
\end{figure*}

\begin{figure*}[th!]
\centering
\includegraphics[width=\textwidth, height=0.17\textheight]{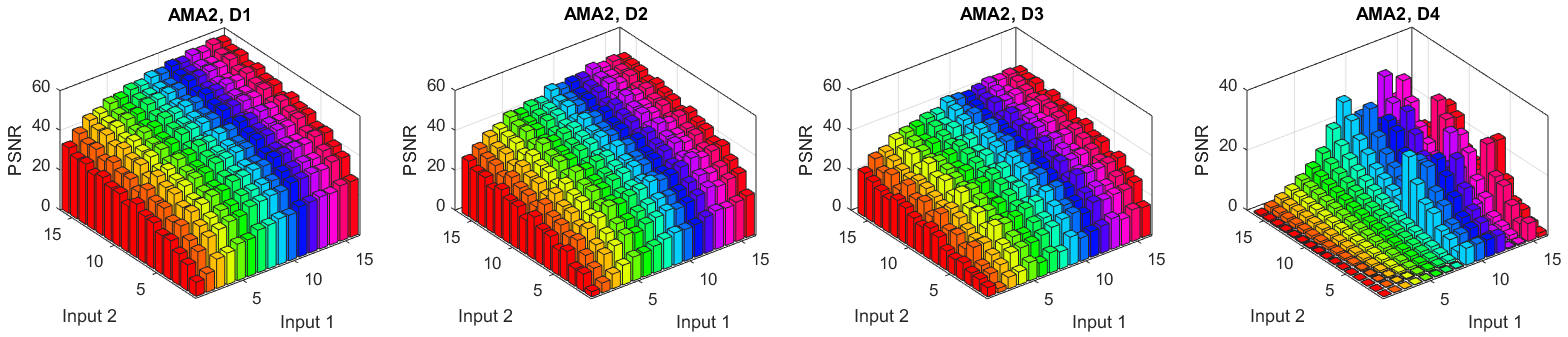} 
\caption{PSNR for approximate multiplier based on \textit{AMA2} type and D1, D2, D3 and D4 approximation degrees}
\label{fig:PSNR_2} 
\vspace{-0.1cm}
\end{figure*}

\begin{figure*}[th!]
\centering
\includegraphics[width=\textwidth, height=0.17\textheight]{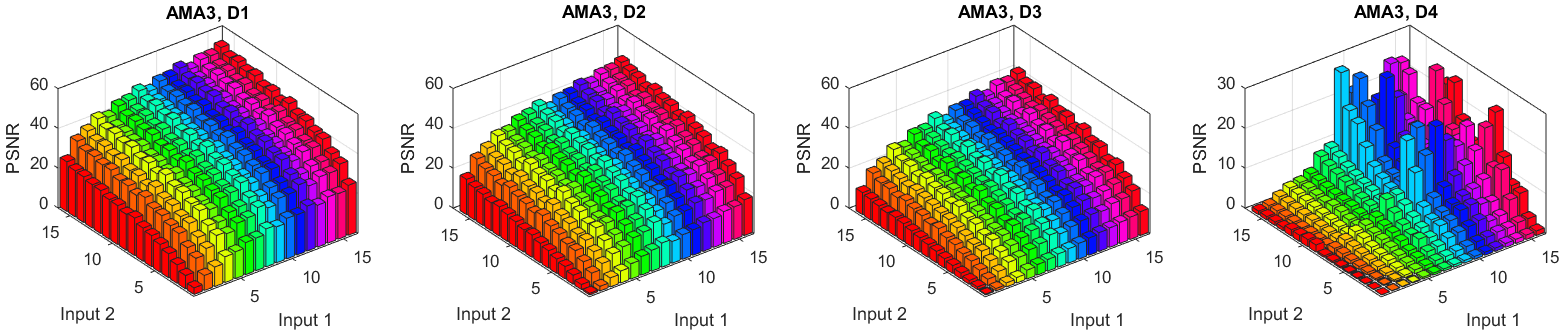} 
\caption{PSNR for approximate multiplier based on \textit{AMA3} type and  D1, D2, D3 and D4 approximation degrees}
\label{fig:PSNR_3} 
\vspace{-0.1cm}
\end{figure*}

\begin{figure*}[th!]
\centering
\includegraphics[width=\textwidth, height=0.17\textheight]{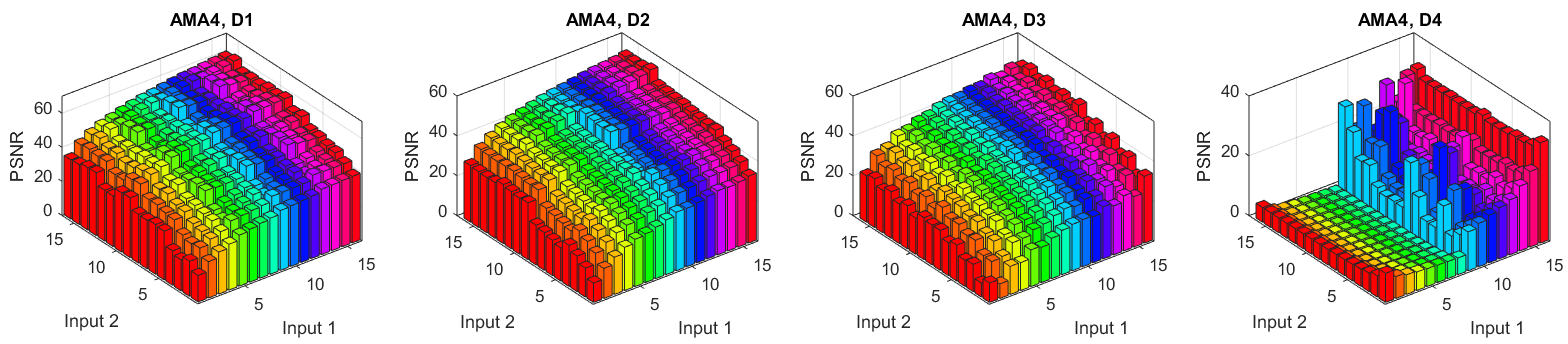} 
\caption{PSNR for approximate multiplier based on \textit{AMA4} type and D1, D2, D3 and D4 approximation degrees}
\label{fig:PSNR_4} 
\vspace{-0.1cm}
\end{figure*}

\begin{figure*}[th!]
\centering
\includegraphics[width=\textwidth, height=0.17\textheight]{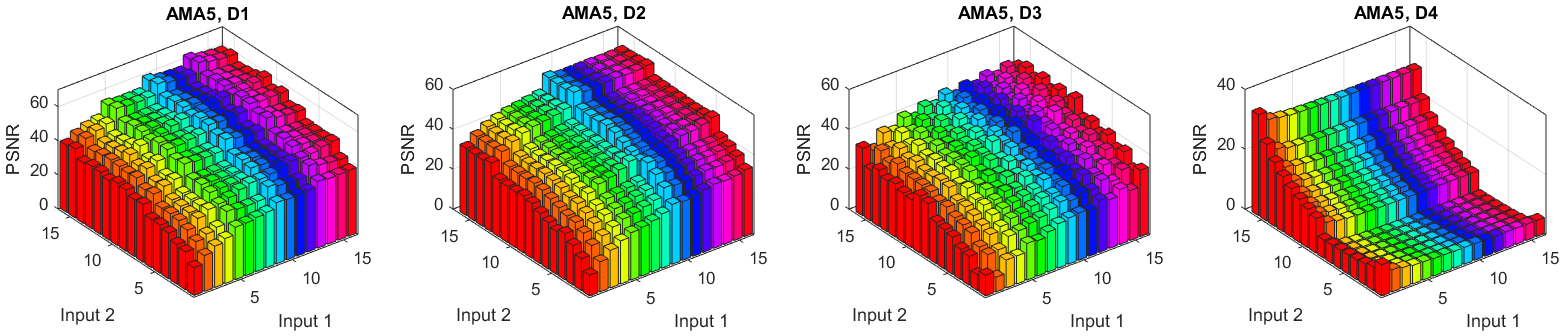} 
\caption{PSNR for approximate multiplier based on \textit{AMA5} type and D1, D2, D3 and D4 approximation degrees}
\label{fig:PSNR_5} 
\vspace{-0.1cm}
\end{figure*}

The Peak Signal-to-Noise Ratio (PSNR), as given in Equation \ref{equ:PSNR}, depends on the Mean Square Error (MSE). Figures \ref{fig:PSNR_1}--\ref{fig:PSNR_5} show the PSNR for different approximate multipliers utilizing \textit{AMA1}--\textit{AMA5} types, respectively. For image processing applications, PSNR is an indication for image quality. Thus, it could be used as TOQ metric while controlling the quality of approximate computing as we proposed in \cite{DATE2019}. A low value of PSNR indicates a low image quality associated with a large MSE. Similar to \cite{PSNR}, we consider PSNR $\geq$ 25 as our threshold for an ``acceptable'' quality.

Figure \ref{fig:PSNR_1} shows the PSNR for designs based on the \textit{AMA1} type. The design based on \textit{D1} has an average value of 39.4dB with a minimum value of 10.7dB. It has 19 out of 256 input combinations with PSNR $<25$dB. Similarly, the design based on \textit{D2} has an average of 32.1dB with a minimum value of 7.7dB, and 54 input combinations with PSNR $<25$dB. The \textit{D3} based design has an average of 25.8dB for the PSNR, with 115 clusters out of the 256 clusters having PSNR $<25$dB. Regarding the \textit{D4} based design, all clusters have a low output quality, with PSNR $<25$dB and a maximum value of 11dB.

The PSNR for designs based on \textit{AMA2} are depicted in Figure \ref{fig:PSNR_2}, where the \textit{D1} based design has an average of 40dB with a minimum value of 7dB. It has 22 input combinations with PSNR $<25$dB. Similarly, the design based on \textit{D2} has an average of 33.2dB with 55 input combinations having PSNR $<25$dB. The \textit{D3} based design has an average of 27.4dB with 99 input combinations having a PSNR $<25$dB. Regarding the \textit{D4} based design, only 7 clusters have a PSNR $>25$dB.

Figure \ref{fig:PSNR_3} depicts a graphical representation for the PSNR of designs based on \textit{AMA3}. The \textit{D1} based design has an average of 34.8dB with 46 input combinations having PSNR $<25$dB, while the \textit{D2} based design has an average of 27.7dB with 92 input combinations having PSNR $<25$dB. The \textit{D3} based design has 151 input combinations, which violate the PSNR, i.e., PSNR $<25$, while the \textit{D4} based design has only 6 clusters with a PSNR $>25$dB. 

The PSNR for designs based on \textit{AMA4} are shown in Figure \ref{fig:PSNR_4}. The design based on \textit{D1} has an average of 45.6dB with a minimum value of 15.5dB. It has only 6 input combinations with PSNR $<25$dB. Similarly, the \textit{D2} based design has an average of 39.1dB with a minimum value of 9.7dB, and 18 input combinations with PSNR $<25$dB. The \textit{D3} based design has an average of 34.3dB for the PSNR, and 43 input combinations with PSNR $<25$dB. Regarding the \textit{D4} based design, 230 clusters have a low output quality, with PSNR $<25$dB and an average value of 10.4dB.

Figure \ref{fig:PSNR_5} shows the PSNR for \textit{AMA5} based designs. The \textit{D1} based design has an average of 46.8dB with a minimum value of 16.3dB, where it has 5 input combinations with PSNR $<25$dB. Similarly, the design based on \textit{D2} has an average of 40.3dB with a minimum value of 10.4dB and 14 input combinations have PSNR $<25$dB. The design with \textit{D3} has an average of 34.6dB for the PSNR, and 33 input combinations have PSNR $<25$dB. Regarding the \textit{D4} based design, 239 clusters have a low output quality with an average of 8.8dB.

Based on the above analysis for the PSNR of 20 different approximate designs, we notice that every design has some input combinations that violate the specified quality constraint, e.g., the PSNR should be at least 25dB. Thus, we should avoid such input-dependent cases with low output quality when requiring a high quality result.

\vspace{-0.4cm}
\section{Conclusion } \label{sec:conclusion}

The emergence of approximate computing has led to the design of interesting designs with low power dissipation, delays and high performance. However, this paradigm requires paying attention to the quality of individual outputs although the average output quality is being acceptable. With this concern, we investigated the relationship between the full range of applicable inputs and their associated approximation errors, where we noticed a strong correlation between them. 

\balance
\newpage
\bibliographystyle{IEEEtran}
\bibliography{main}
\end{document}